\documentstyle[amsfonts,12pt,latexsym]{article}

\def\beqa{\begin{eqnarray}}
\def\eeqa{\end{eqnarray}}
\def\beq{\begin{equation}}
\def\eeq{\end{equation}}
\newcommand{\ba}{\begin{array}}
\newcommand{\ea}{\end{array}}
\begin{document}

\begin{titlepage}
\begin{flushright}
IFUM-655-FT\\
CAMS/00-01\\
hep-th/0001131
\end{flushright}
\vspace{.3cm}
\begin{center}
\renewcommand{\thefootnote}{\fnsymbol{footnote}}
{\Large \bf Supersymmetry of Black Strings in $D=5$
Gauged Supergravities}
\vfill
{\large \bf {D.~Klemm$^1$\footnote{email: dietmar.klemm@mi.infn.it} and
W.~A.~Sabra$^2$\footnote{email: ws00@aub.edu.lb}}}\\
\renewcommand{\thefootnote}{\arabic{footnote}}
\setcounter{footnote}{0}
\vfill
{\small
$^1$ Dipartimento di Fisica dell'Universit\`a di Milano
and\\ INFN, Sezione di Milano,
Via Celoria 16, 20133 Milano, Italy.\\
\vspace*{0.4cm}
$^2$ Center for Advanced Mathematical Sciences (CAMS)\\
and\\
Physics Department, American University of Beirut, Lebanon.}
\end{center}
\vfill
\begin{center}
{\bf Abstract}
\end{center}

Supersymmetry of five dimensional string solutions is examined in the
context of gauged $D=5$, $N=2$ supergravity coupled to abelian vector
multiplets. We find magnetic black strings preserving one quarter of
supersymmetry and approaching the half-supersymmetric product space
$AdS_3\times H^2$ near the event horizon. The solutions thus exhibit the
phenomenon of supersymmetry enhancement near the horizon, like in the
cases of ungauged supergravity theories, where the near horizon limit is
fully supersymmetric. Finally, product space compactifications are studied
in detail, and it is shown that only for negative curvature (hyperbolic)
internal spaces, some amount of supersymmetry can be preserved. Among other
solutions, we find that the extremal rotating BTZ black hole tensored by
$H^2$ preserves one quarter of supersymmetry.
\end{titlepage}

\bigskip

\bigskip

\section{Introduction}

The conjectured equivalence between string theory on anti-de~Sitter (AdS)
spaces (times some compact manifold) and certain superconformal gauge
theories living on the boundary of AdS \cite{ads} has led to an increasing
interest in black objects in asymptotically anti-de~Sitter spaces. On one
hand, this interest is based on the fact that the classical supergravity
solution can furnish important information on the dual gauge theory in the
large $N$ limit, $N$ being the rank of the gauge group. An example of this
is the Hawking-Page phase transition \cite{hawkpage} from thermal AdS space
to the Schwarz\-schild-AdS black hole, which was later reconsidered by
Witten in the spirit of the AdS/CFT correspondence \cite{witten}. There it
was observed that it can be interpreted as a transition from a
low-temperature confining to a high temperature deconfining phase in the
dual field theory.\\
On the other hand, the proposed AdS/CFT equivalence opens the possibility to
a microscopic understanding of the Bekenstein-Hawking entropy of
asymptotically anti-de~Sitter black holes. This route was pioneered by
Strominger \cite{strominger}, who used the central charge of the AdS$_{3}$
asymptotic symmetry algebra \cite{brown} to count the microstates giving
rise to the BTZ black hole entropy. \newline
Of particular interest in this context are black objects in AdS space which
preserve some fraction of supersymmetry. On the CFT side, these supergravity
vacua could correspond to an expansion around non-zero vacuum expectation
values of certain operators. Supersymmetry of Reissner-Nordstr\"{o}m-AdS
black holes in four dimensions was first studied by Romans in the context of
$N=2$ gauged supergravity \cite{romans}. These considerations have been
extended later in various
directions \cite{london,perry,bcs1,ck,klemm,bcs2,duff,liu,sabra,chamsabra}.
One common feature of all results is the appearance of naked singularities
in the BPS limit\footnote{It is worth pointing out, however, that by
including rotation \cite{perry,ck}, or by allowing for different event
horizon geometries \cite{ck,klemm}, one
can get genuine BPS black holes in anti-de~Sitter space.}. This means that
the theory on the bulk side is ill-defined in the limit of small distances,
and stringy corrections have to be taken into account.\newline
Although many string- and brane solutions in ungauged supergravity theories
are known, very little is known on the corresponding objects in the gauged
case. To remedy this will be the main purpose of the present paper. In
particular, we will derive supersymmetric black string solutions with
various topologies in five dimensional $N=2$ supergravity theories coupled
to vector multiplets \cite{gst2}. The theory of ungauged five-dimensional
$N=2$ supergravity coupled to abelian vector supermultiplets can be obtained
by compactifying eleven-dimensional supergravity, the low-energy theory of
M-theory, on a Calabi-Yau three-fold \cite{cy}. Gauged supergravity theories
are obtained by gauging a subgroup of the R-symmetry group, the automorphism
group of the supersymmetry algebra. The gauged $D=5$, $N=2$ supergravity
theories are obtained by gauging the $U(1)$ subgroup of the $SU(2)$
automorphism group of the superalgebra \cite{gst2}. The Lagrangian of the
theory is obtained by introducing a linear combination of the abelian vector
fields already present in the ungauged theory,
i.~e.~$A_{\mu }=V_{I}A_{\mu}^{I}$, with a coupling constant $g$. The coupling
of the Fermi-fields to the $U(1)$ vector field breaks supersymmetry, and
therefore gauge-invariant $g$-dependent terms must be introduced in order to
preserve $N=2$ supersymmetry. In a bosonic background, this amounts to the
addition of a $g^2$-dependent scalar potential $V$ \cite{bcs1,gst2}.\newline
Our work in this paper is organized as follows. Section \ref{sugra} contains
a brief review of $D=5$, $N=2$ gauged supergravity. In \ref{susystrsol}, the
supersymmetric string solutions are derived, and their near horizon limit is
studied. In section \ref{general}, general supersymmetric product space
compactifications of $D=5$, $N=2$ gauged supergravity are considered.
Finally, our results are summarized and discussed in \ref{disc}.

\section{$D=5$, $N=2$ Gauged Supergravity}

\label{sugra}

The bosonic part of the gauged supersymmetric $N=2$ Lagrangian which
describes the coupling of vector multiplets to supergravity is given by
\begin{eqnarray}
e^{-1}{\cal {L}} &=&\frac{1}{2}R+g^{2}V-{\frac{1}{4}}G_{IJ}F_{\mu \nu
}{}^{I}F^{\mu \nu J}-\frac{1}{2}{\cal G}_{ij}\partial _{\mu }\phi
^{i}\partial ^{\mu }\phi ^{j}  \nonumber \\
&&+{\frac{e^{-1}}{48}}\epsilon ^{\mu \nu \rho \sigma \lambda }C_{IJK}F_{\mu
\nu }^{I}F_{\rho \sigma }^{J}A_{\lambda }^{K},  \label{action}
\end{eqnarray}
where $\mu ,\nu $ are spacetime indices, $R$ is the scalar curvature, $%
F_{\mu \nu }^{I}$ denote the abelian field-strength tensors, and $e=\sqrt{-g}
$ is the determinant of the F\"{u}nfbein $e_{\mu }^{a}$. The scalar
potential $V$ is given by 
\[
V(X)=V_{I}V_{J}\left( 6X^{I}X^{J}-{\frac{9}{2}}{\cal G}^{ij}\partial
_{i}X^{I}\partial _{j}X^{J}\right) , 
\]
where $X^{I}$ represent the real scalar fields satisfying the condition $%
{\cal V}={\frac{1}{6}}C_{IJK}X^{I}X^{J}X^{K}=1.$ The physical quantities in (%
\ref{action}) can all be expressed in terms of the homogeneous cubic
polynomial ${\cal V}$ which defines a ``very special geometry''\cite{dewit}.
We also have the relations 
\begin{eqnarray}
G_{IJ} &=&-{\frac{1}{2}}\partial _{I}\partial _{J}\log {\cal V}\Big|_{{\cal V%
}=1},  \nonumber \\
{\cal G}_{ij} &=&\partial _{i}X^{I}\partial _{j}X^{J}G_{IJ}\Big|_{{\cal V}%
=1},
\end{eqnarray}
where $\partial _{i}$ and $\partial _{I}$ refer, respectively, to a partial
derivative with respect to the scalar field $\phi ^{i}$ and $%
X^{I}=X^{I}(\phi ^{i})$.\newline
Note that for Calabi-Yau compactification of M-theory, ${\cal V}$ is the
intersection form, $X^{I}$ and $X_{I}=\frac{1}{6}C_{IJK}X^{J}X^{K}$
correspond to the size of the two- and four-cycles and $C_{IJK}$ are the
intersection numbers of the Calabi-Yau threefold.\newline
The supersymmetry transformations of the gravitino $\psi _{\mu }$ and the
gauginos $\lambda _{i}$ in a bosonic background read \cite{bcs1} 
\begin{eqnarray}
\delta \psi _{\mu } &=&\left( {\cal D}_{\mu }+\frac{i}{8}X_{I}(\Gamma _{\mu
}{}^{\nu \rho }-4\delta _{\mu }{}^{\nu }\Gamma ^{\rho })F_{\nu \rho }{}^{I}+%
\frac{1}{2}g\Gamma _{\mu }X^{I}V_{I}-\frac{3i}{2}gV_{I}A_{\mu }^{I}\right)
\epsilon ,  \label{stgrav} \\
\delta \lambda _{i} &=&\left( {\frac{3}{8}}\Gamma ^{\mu \nu }F_{\mu \nu
}^{I}\partial _{i}X_{I}-{\frac{i}{2}}{\cal G}_{ij}\Gamma ^{\mu }\partial
_{\mu }\phi ^{j}+\frac{3i}{2}gV_{I}\partial _{i}X^{I}\right) \epsilon ,
\label{stgaug}
\end{eqnarray}
where $\epsilon $ is the supersymmetry parameter and ${\cal {D}_{\mu }}$ is
the covariant derivative\footnote{%
We use the metric $\eta ^{ab}=(-,+,+,+,+)$, $\{{\Gamma ^{a},\Gamma ^{b}}%
\}=2\eta ^{ab}$, ${\cal {D}}_{\mu }=\partial _{\mu }+{\ \frac{1}{4}}\omega
_{\mu ab}\Gamma ^{ab}$, $\omega _{\mu ab}$ is the spin connection, and $%
\Gamma ^{{a}_{1}{a}_{2}\cdots {a_{n}}}={\frac{1}{n!}}\Gamma ^{\lbrack {a_{1}}%
}\Gamma ^{{a_{2}}}\cdots \Gamma ^{{a_{n}}]}$.}.

\section{Supersymmetric String Solutions}

\label{susystrsol}

As a general ansatz for the supersymmetric solutions we consider metrics of
the form
\begin{equation}
ds^{2}=-e^{2V}dt^{2}+e^{2T}dz^{2}+e^{2U}dr^{2}+F(r)^{2}d\sigma ^{2},
\label{metric}
\end{equation}
where $V,T,U$ are functions of $r$ only, and we consider either $F(r)=r$ or $%
F(r)=R=$ constant. $d\sigma ^{2}$ denotes the metric of a two-manifold $%
{\cal S}$ of constant Gaussian curvature $k$. Without loss of generality we
restrict ourselves to the cases $k=0,\pm 1$. Clearly ${\cal S}$ is a
quotient space of the universal coverings $S^{2}$ ($k=1$), $H^{2}$ ($k=-1$)
or $E^{2}$ ($k=0$). Explicitly, we choose 
\begin{equation}
d\sigma ^{2}=d\theta ^{2}+f(\theta )^{2}d\phi ^{2},
\end{equation}
where 
\begin{equation}
f(\theta )=\left\{ 
\begin{array}{l@{\quad,\quad}l}
\sin \theta & k=1, \\ 
1 & k=0, \\ 
\sinh \theta & k=-1.
\end{array}
\right.
\end{equation}
The case $k=1$, $F=r$ has been considered in \cite{chamsabra}. There, a
spherically symmetric magnetic string solution was obtained, which contains
a naked singularity.\newline
For the metric (\ref{metric}), the f\"{u}nfbein and its inverse can be
chosen as
\begin{eqnarray}
e_{t}^{0} &=&e^{V},\qquad e_{z}^{1}=e^{T},\qquad e_{r}^{2}=e^{U},\qquad
e_{\theta }^{3}=F,\qquad e_{\phi }^{4}=Ff,  \nonumber \\
e_{0}^{t} &=&e^{-V},\qquad e_{1}^{z}=e^{-T},\qquad e_{2}^{r}=e^{-U},\qquad
e_{3}^{\theta }=\frac{1}{F},\qquad e_{4}^{\phi }=\frac{1}{Ff}.
\end{eqnarray}
The nonvanishing components of the spin connection are given by 
\begin{eqnarray}
\omega _{t}^{02} &=&V^{\prime }e^{V-U},  \nonumber \\
\omega _{z}^{12} &=&T^{\prime }e^{T-U},  \nonumber \\
\omega _{\theta }^{23} &=&-F^{\prime }e^{-U},  \nonumber \\
\omega _{\phi }^{24} &=&-F^{\prime }fe^{-U},  \nonumber \\
\omega _{\phi }^{34} &=&-f^{\prime }.  \label{spinconn}
\end{eqnarray}
In five dimensions, strings can carry magnetic charges under the one-form
potentials $A^{I}$, so we assume that the gauge fields have only a magnetic
part, i.~e. 
\begin{equation}
F_{\theta \phi }^{I}=kq^{I}f(\theta ),\qquad A_{\phi }^{I}=kq^{I}\int
f(\theta )d\theta .  \label{magnfield}
\end{equation}
Furthermore, we are looking for solutions with constant moduli $X^{I}$,
which are chosen to minimize the magnetic central charge $Z=q^{I}X_{I}$, as
in the case of the double extreme solutions in the ungauged theory \cite
{feka,magnetic}. This means that one has 
\begin{equation}
\partial _{i}(Z)=\partial _{i}(q^{I}X_{I})=\frac{1}{3}C_{IJK}X^{I}\partial
_{i}(X^{J})q^{K}=0.  \label{mincc}
\end{equation}
Moreover, we make the choice 
\begin{equation}
X^{I}V_{I}=1.  \label{choice}
\end{equation}
Using (\ref{mincc}) and (\ref{choice}), the gaugino transformations (\ref
{stgaug}) can be easily seen to vanish identically.\newline
Plugging the spin connection (\ref{spinconn}) and the magnetic fields (\ref
{magnfield}) into (\ref{stgrav}), we obtain for the supersymmetry variation
of the gravitino 
\begin{eqnarray}
\delta \psi _{t} &=&\left( \partial _{t}+\frac{1}{2}V^{\prime }e^{V-U}\Gamma
_{02}+\frac{ik}{4}Z\frac{e^{V}}{F^{2}}\Gamma _{034}+\frac{g}{2}e^{V}\Gamma
_{0}\right) \epsilon ,  \nonumber \\
\delta \psi _{z} &=&\left( \partial _{z}+\frac{1}{2}T^{\prime }e^{T-U}\Gamma
_{12}+\frac{ik}{4}Z\frac{e^{T}}{F^{2}}\Gamma _{134}+\frac{g}{2}e^{T}\Gamma
_{1}\right) \epsilon ,  \nonumber \\
\delta \psi _{r} &=&\left( \partial _{r}+{\frac{ike^{U}}{4}}\frac{Z}{F^{2}}%
\Gamma _{234}+\frac{g}{2}e^{U}\Gamma _{2}\right) \epsilon ,  \label{gravvar}
\\
\delta \psi _{\theta } &=&\left( \partial _{\theta }-\frac{1}{2}F^{\prime
}e^{-U}\Gamma _{23}-\frac{ikZ}{2F}\Gamma _{4}+\frac{g}{2}F\Gamma _{3}\right)
\epsilon ,  \nonumber \\
\delta \psi _{\phi } &=&\left( \partial _{\phi }-\frac{1}{2}F^{\prime
}fe^{-U}\Gamma _{24}-\frac{1}{2}f^{\prime }\Gamma _{34}+\frac{ikZf}{2F}%
\Gamma _{3}+\frac{g}{2}Ff\Gamma _{4}-\frac{3i}{2}gV_{I}q^{I}k\int fd\theta
\right) \epsilon .  \nonumber
\end{eqnarray}
In what follows, we will consider three different cases.

\subsection{Solutions With Flat Transverse Space}

Let us first consider the case $k=0$, $F=r$, i.~e.~flat transverse space.
Our choice (\ref{magnfield}) implies that we have vanishing gauge fields for 
$k=0$. The vanishing of the gravitino supersymmetry transformations (\ref
{gravvar}) then yields the Killing spinor equations 
\begin{eqnarray}
\left( \partial _{t}+\frac{1}{2}e^{V}\Gamma _{0}(V^{\prime }e^{-U}\Gamma
_{2}+g)\right) \epsilon &=&0,  \nonumber \\
\left( \partial _{z}+\frac{1}{2}e^{T}\Gamma _{1}(T^{\prime }e^{-U}\Gamma
_{2}+g)\right) \epsilon &=&0,  \nonumber \\
\left( \partial _{r}+\frac{g}{2}e^{U}\Gamma _{2}\right) \epsilon &=&0, 
\nonumber \\
\left( \partial _{\theta }+\frac{1}{2}\Gamma _{3}(e^{-U}\Gamma
_{2}+gr)\right) \epsilon &=&0,  \nonumber \\
\left( \partial _{\phi }+\frac{1}{2}\Gamma _{4}(e^{-U}\Gamma _{2}+gr)\right)
\epsilon &=&0.  \label{remsys}
\end{eqnarray}
From the integrability conditions of these equations one gets
\begin{equation}
e^{V}=e^{T}=e^{-U}=gr.  \label{VTU}
\end{equation}
Plugging these results into (\ref{remsys}) and introducing the projectors $%
P_{\pm }=\frac{1}{2}(1\pm \Gamma _{2})$, we obtain for the Killing spinors 
\begin{equation}
\epsilon =r^{-\frac{1}{2}}\epsilon _{+}^{0}-gr^{\frac{1}{2}}(g\Gamma
_{0}t+g\Gamma _{1}z+\Gamma _{3}\theta +\Gamma _{4}\phi )\epsilon _{+}^{0}+r^{%
\frac{1}{2}}\epsilon _{-}^{0},  \label{killspin}
\end{equation}
where $\epsilon _{\pm }^{0}$ are constant spinors satisfying $P_{\mp
}\epsilon _{\pm }=0$. From (\ref{VTU}) it is clear that the solution we
found is locally $AdS_{5}$ (written in horospherical coordinates). (\ref
{killspin}) tells us that this spacetime, as it should be, is fully
supersymmetric. However, one may wish to compactify the $(\theta ,\phi )$
sector to a cylinder or a torus, considering thus a quotient space of $%
AdS_{5}$. In this case, the surviving Killing spinors are those which
respect the identifications performed in the $(\theta ,\phi )$ sector. These
are 
\begin{equation}
\epsilon =r^{\frac{1}{2}}\epsilon _{-}^{0},
\end{equation}
so that the considered $AdS_{5}$ quotient space preserves half of the
supersymmetries.\newline
Note that the above supersymmetric string solution is a limiting case
of a family of nonextremal black strings, whose metric is given by
\begin{equation}
ds^{2}=-e^{2V}dt^{2}+r^{2}dz^{2}+e^{-2V}dr^{2}+r^{2}(d\theta ^{2}+d\phi
^{2}),  \label{nonextr}
\end{equation}
where 
\begin{equation}
e^{2V}=-\frac{m}{r^{2}}+g^{2}r^{2},
\end{equation}
$m$ denoting an integration constant related to the mass of the black
string. Considering $z$ as a coordinate of transverse space, (\ref{nonextr})
clearly can also be interpreted as a black {\it hole}, it is the solution
found in \cite{birmingh}.

\subsection{Hyperbolic Transverse Space}

We turn now to the more interesting case of hyperbolic transverse space,
i.~e.~$k=-1$, $F=r$. As supersymmetry breaking conditions we take 
\begin{eqnarray}
\Gamma_3\Gamma_4\epsilon &=& i\epsilon,  \nonumber \\
\Gamma_2\epsilon &=& -\epsilon.
\end{eqnarray}
Then, the transformations (\ref{stgrav}) reduce to 
\begin{eqnarray}
\delta\psi_t &=& \left(\partial_t - \frac 12(V^{\prime }e^{V-U} - Z \frac{e^V%
}{2r^2} - ge^V)\Gamma_0\right)\epsilon,  \nonumber \\
\delta\psi_z &=& \left(\partial_z - \frac 12(T^{\prime }e^{T-U}-Z \frac{e^T}{%
2r^2} - ge^T)\Gamma_1\right)\epsilon,  \nonumber \\
\delta \psi_r &=& \left(\partial_r - \frac{e^U}{2}(\frac{Z}{2r^2} + g)
\right)\epsilon, \\
\delta\psi_{\theta} &=& \left(\partial_{\theta} - \frac 12(e^{-U} + \frac Zr
- gr)\Gamma_3\right)\epsilon,  \nonumber \\
\delta\psi_{\phi} &=& \left(\partial_{\phi} - \frac i2
(1-3gV_Iq^I)\cosh\theta - \frac 12(e^{-U} + \frac Zr - gr)\sinh\theta
\Gamma_4\right)\epsilon.  \nonumber
\end{eqnarray}

The vanishing of the above equations implies the following conditions on the
supersymmetry spinor $\epsilon$, 
\begin{eqnarray}
\partial_t\epsilon &=& 0,  \nonumber \\
\partial_z\epsilon &=& 0,  \nonumber \\
\partial_{\theta}\epsilon &=& 0,  \nonumber \\
\partial_{\phi}\epsilon &=& 0,  \nonumber \\
3gq^IV_I &=& 1,  \nonumber \\
-e^{-U} - \frac Zr + gr &=& 0,  \nonumber \\
-e^{-U}T^{\prime } + \frac{Z}{2r^2} + g &=& 0,  \nonumber \\
-e^{-U}V^{\prime } + \frac{Z}{2r^2} + g &=& 0.  \label{set}
\end{eqnarray}

\bigskip The last two equations in (\ref{set}) imply that one should set $%
T=V $. From the sixth equation of (\ref{set}), one immediately obtains 
\begin{equation}
e^{-U} = -\frac Zr + gr.
\end{equation}
Using the last equation of (\ref{set}), we obtain a differential equation
for $V$, 
\begin{equation}
V^{\prime}e^{-U} = g + \frac{Z}{2r^2}.  \label{de}
\end{equation}
The above differential equation can be easily solved by noticing that it can
be rewritten in the form 
\begin{equation}
\frac{dV}{dr} = \frac{d}{dr}\log(e^{-U}) - \frac 14\frac{d}{dr} \log(\frac{%
e^{-U}}{gr}).  \label{diffequ}
\end{equation}
(Recall that the central charge $Z$ takes a constant value to be
determined). (\ref{diffequ}) yields the following solution for $V$, 
\begin{equation}
e^{V}=e^{-3{\frac{U}{4}}}(gr)^{{\frac{1}{4}}}.
\end{equation}

Let us now return to the minimization condition (\ref{mincc}) of the
magnetic central charge. It implies that the critical values of $X^{I}$ and
its dual are given by 
\begin{equation}
X^{I}=\frac{q^{I}}{Z},\qquad X_{I}=\frac{1}{6}\frac{C_{IJK}q^{J}q^{K}}{Z^{2}}%
,
\end{equation}
and thus the critical value of the magnetic central charge is
\begin{equation}
Z = \left(\frac{1}{6}C_{IJK}q^{I}q^{J}q^{K}\right)^{\frac 13}.
\end{equation}
Using the conditions $X^{I}V_{I}=1$ and the fifth relation of (\ref{set}),
one obtains a generalized Dirac quantization condition 
\begin{equation}
\left(\frac{1}{6}C_{IJK}q^{I}q^{J}q^{K}\right)^{\frac 13}=\frac{1}{3g}.
\end{equation}
For the case of pure supergravity where only the graviphoton charge $q^{0}$
is present, one obtains $q^0 = 1/(3g)$. A similar condition was obtained in 
\cite{romans,ck}.

\bigskip To summarize, the BPS magnetic black string solution to $D=5$, $N=2$
gauged supergravity coupled to vector multiplets is given by 
\begin{eqnarray}
ds^2 &=& (gr)^{\frac 12}e^{-\frac 32 U}(-dt^2 + dz^2) + e^{2U}dr^2 +
r^2\left(d\theta^2 + \sinh^2\theta d\phi^2\right),  \nonumber \\
e^{-U} &=& -\frac{1}{3gr} + gr,  \label{magstrsol}
\end{eqnarray}
while the gauge fields and the scalars are 
\begin{eqnarray}
A_{\phi}^I &=& -q^I\cosh\theta,  \nonumber \\
X^I &=& 3gq^I.
\end{eqnarray}

\bigskip The Killing spinor is independent of the angular variables, and its
radial dependence is obtained by solving for its radial differential
equation, which reads 
\begin{equation}
\left(\partial_r - \frac{e^{U}}{2}(g + \frac{1}{6gr^2})\right)\epsilon = 0.
\end{equation}
Using the relation (\ref{de}), the above differential equation can be
written in the simple form 
\begin{equation}
\left(\partial_r - \frac 12 V^{\prime}\right)\epsilon = 0,
\end{equation}
and thus we get 
\begin{equation}
\epsilon(r) = e^{\frac 12 V}\epsilon_0,
\end{equation}
where $\epsilon_0$ is a constant spinor satisfying the constraints 
\begin{equation}
\Gamma_3\Gamma_4\epsilon_0 = i\epsilon_0, \qquad \Gamma_2\epsilon_0 =
-\epsilon_0.
\end{equation}
As the Killing spinors do not depend on the coordinates $\theta,\phi$ of the
transverse hyperbolic space, one could also compactify the $H^2$ to a
Riemann surface ${\cal S}_n$ of genus $n$, and the resulting solution would
still preserve one quarter of supersymmetry.\newline
Whereas the spherical BPS magnetic string found in \cite{chamsabra} contains
a naked singularity, the hyperbolic black string (\ref{magstrsol}) has an
event horizon at $r = r_+ = 1/(g\sqrt 3)$. This is analogous to the $AdS_4$
case, where BPS magnetic black holes with hyperbolic event horizons have
been found \cite{ck}, whereas for spherical topology one gets supersymmetric
naked singularities \cite{romans}. \newline
Note that the black strings (\ref{magstrsol}) are solitonic objects in the
sense that the limit $g\to 0$ (we recall that $g$ is the coupling constant
of the gauged theory, coupling the $U(1)$ vector fields to the fermions)
does not exist.\newline
In the near horizon region, (\ref{magstrsol}) reduces to the product
manifold $AdS_3 \times H^2$. This is easily seen by introducing the new
radial coordinate $\rho = (r-r_+)^{1/4}$. In the next subsection, we will
see that in the near horizon limit, supersymmetry is enhanced.

\subsection{Hyperbolic Transverse Space and Constant Warping Function}

Let us now consider the case $k=-1$ and $F=R$, where $R$ is a constant. We
also choose $T=V$. As supersymmetry breaking condition we take 
\begin{equation}
\Gamma_3\Gamma_4\epsilon = i\epsilon.
\end{equation}
The Killing spinor equations following from (\ref{stgrav}) are then 
\begin{eqnarray}
\delta\psi_t &=& \left(\partial_t + \frac 12V^{\prime}e^{V-U}\Gamma_{02} +
\frac 14 Z\frac{e^V} {R^2}\Gamma_0 + \frac g2 e^V \Gamma_0\right)\epsilon =
0,  \nonumber \\
\delta\psi_z &=& \left(\partial_z + \frac 12V^{\prime}e^{V-U}\Gamma_{12} +
\frac 14 Z\frac{e^V} {R^2}\Gamma_1 + \frac g2 e^V \Gamma_1\right)\epsilon =
0,  \nonumber \\
\delta\psi_r &=& \left(\partial_r + \frac 14 Z \frac{e^U}{R^2}\Gamma_2 +
\frac g2 e^U\Gamma_2 \right)\epsilon = 0,  \label{kspeq} \\
\delta\psi_{\theta} &=& \left(\partial_{\theta} - \frac{Z}{2R}\Gamma_3 +
\frac g2 R\Gamma_3\right) \epsilon = 0,  \nonumber \\
\delta\psi_{\phi} &=& \left(\partial_{\phi} - \frac i2 \cosh\theta - \frac{Z%
}{2R}\sinh\theta\Gamma_4 + \frac{gR}{2}\sinh\theta\Gamma_4 + \frac{3i}{2}%
gV_Iq^I\cosh\theta\right) \epsilon = 0.  \nonumber
\end{eqnarray}
The integrability conditions for these equations imply that 
\begin{equation}
3gV_Iq^I =1,  \label{3gVq}
\end{equation}
and that the central charge is related to the compactification radius by $%
Z=gR^2$. Furthermore, one obtains 
\begin{equation}
\partial_r(V^{\prime}e^{V-U}) = \frac 94 g^2e^{V+U}, \qquad V^{\prime}e^{-U}
= \frac 32 g.
\end{equation}
Defining a new radial coordinate $\rho$ by $g\rho = e^V$, one immediately
sees that the three-dimensional part of spacetime is $AdS_3$ in
horospherical coordinates. We have thus obtained a supersymmetric product
space $AdS_3 \times H^2$, i.~e.~the near-horizon geometry of (\ref{magstrsol}%
). Plugging the relations (\ref{3gVq}) and $Z=gR^2$ into (\ref{kspeq}), one
obtains that the Killing spinors are independent of $\theta,\phi$. The
remaining system is solved by 
\begin{equation}
\epsilon = e^{-\frac 12 V}\epsilon^0_+ - \frac 32 g e^{\frac 12 V}(\Gamma_0
t + \Gamma_1 z)\epsilon^0_+ + e^{\frac 12 V}\epsilon^0_-,
\end{equation}
where $\epsilon^0_{\pm}$ are constant spinors satisfying 
\begin{equation}
(1\mp\Gamma_2)\epsilon^0_{\pm} = 0, \qquad \Gamma_3\Gamma_4\epsilon^0_{\pm}
= i\epsilon^0_{\pm}.
\end{equation}
The product space $AdS_3 \times H^2$ is thus one half supersymmetric. This
supersymmetry enhancement near the horizon of the BPS black string is
analogous to the case of ungauged supergravity theories, where usually in
the near-horizon limit supersymmetry is fully restored.

\section{General Product Space Compactifications}

\label{general}

In this section we consider general product space compactifications of
gauged $D=5$, $N=2$ supergravity coupled to vector multiplets. Spacetime is
assumed to be a product $M_{3}\times M_{2}$, where, as above, $M_{2}$
denotes a two-manifold of constant curvature. We are interested in the
conditions imposed by supersymmetry on $M_{2,3}$. To this end, we perform a $%
3+2$ decomposition of the gamma matrices\footnote{%
See the appendix of \cite{pope} for a nice summary of gamma matrix
decomposition in Kaluza-Klein compactifications.} in the following way 
\begin{equation}
\Gamma ^{a}=(\Gamma ^{\hat{\alpha}},\Gamma ^{\hat{\imath}})=(\gamma ^{\hat{%
\alpha}}\otimes \sigma ^{3},1\otimes \Sigma ^{\hat{\imath}}),
\end{equation}
where early Greek letters $\alpha ,\beta ,\ldots $ are $M_{3}$ spacetime
indices, and $i,j,\ldots $ are $M_{2}$ spacetime indices. The hatted indices
refer to the corresponding tangent spaces. The $\gamma ^{\hat{\alpha}}$ and $%
\Sigma ^{\hat{\imath}}$ denote Dirac matrices in three and two dimensions
respectively. To be concrete, we make the choice $\Sigma ^{1}=\sigma ^{2}$, $%
\Sigma ^{2}=\sigma ^{1}$, where the Pauli matrices are chosen to be 
\begin{equation}
\sigma ^{1}=\left( 
\begin{array}{cc}
0 & 1 \\ 
1 & 0
\end{array}
\right) ,\qquad \sigma ^{2}=\left( 
\begin{array}{cc}
0 & -i \\ 
i & 0
\end{array}
\right) ,\qquad \sigma ^{3}=\left( 
\begin{array}{cc}
1 & 0 \\ 
0 & -1
\end{array}
\right) .
\end{equation}
The supersymmetry parameter $\epsilon $ in five dimensions is decomposed as $%
\epsilon =\eta \otimes \chi $. Note that $\sigma ^{3}$ plays the role of a
chirality operator for the spinors $\chi $ in two dimensions. Some useful
relations needed below are 
\begin{equation}
\Gamma ^{\hat{\alpha}\hat{\beta}}=\gamma ^{\hat{\alpha}\hat{\beta}}\otimes
1,\qquad \Gamma ^{\hat{\imath}\hat{\jmath}}=1\otimes \Sigma ^{\hat{\imath}%
\hat{\jmath}},\qquad \Gamma ^{\hat{\alpha}}\Gamma ^{\hat{\imath}}\Gamma ^{%
\hat{\jmath}}=\gamma ^{\hat{\alpha}}\otimes \sigma ^{3}\Sigma ^{\hat{\imath}%
}\Sigma ^{\hat{\jmath}}.
\end{equation}
For the field strength of the abelian vectors we make the ansatz 
\begin{equation}
F^{I}=q^{I}\epsilon ,
\end{equation}
where $\epsilon $ denotes the volume form on $M_{2}$\footnote{%
We apologize for using the same symbol for the volume form and the
supersymmetry parameter, but the meaning should be clear from the context.}.
The gaugino variation (\ref{stgaug}) vanishes as before, provided that (\ref
{mincc}) and $V_{I}X^{I}=1$ are satisfied. (Note that we still assume the
moduli $X^{I}$ to be constant). Inserting the decomposition of the Dirac
matrices and the supersymmetry parameter, as well as the ansatz for the
field strength into the gravitino variation (\ref{stgrav}) yields the
relations 
\begin{eqnarray}
\lefteqn{\left( \partial _{\alpha }+\frac{1}{4}\omega _{\alpha }^{\hat{\alpha%
}\hat{\beta}}\gamma _{\hat{\alpha}\hat{\beta}}+\frac{1}{4}X_{I}q^{I}\gamma
_{\alpha }+\frac{g}{2}\gamma _{\alpha }\right) \eta \otimes \chi _{+}} 
\nonumber \\
&&+\left( \partial _{\alpha }+\frac{1}{4}\omega _{\alpha }^{\hat{\alpha}\hat{%
\beta}}\gamma _{\hat{\alpha}\hat{\beta}}+\frac{1}{4}X_{I}q^{I}\gamma
_{\alpha }-\frac{g}{2}\gamma _{\alpha }\right) \eta \otimes \chi _{-}=0
\label{etaeq}
\end{eqnarray}
and 
\begin{equation}
\left( \partial _{i}+\frac{1}{4}\omega _{i}^{\hat{\imath}\hat{\jmath}}\Sigma
_{\hat{\imath}\hat{\jmath}}-\frac{i}{2}X_{I}q^{I}\epsilon _{ij}\Sigma ^{j}+%
\frac{g}{2}\Sigma _{i}-\frac{3i}{2}gV_{I}A_{i}^{I}\right) \chi =0,
\label{chieq}
\end{equation}
where $\chi _{\pm }=\frac{1}{2}(1\pm \sigma ^{3})\chi $ denote the chirality
projections of the two-dimensional spinor $\chi $. From (\ref{etaeq}) we get
that either $\chi _{-}$ and the bracketed expression in the first line have
to vanish, or $\chi _{+}$ and the term in brackets in the second line are
zero. Without loss of generality, we assume the first possibility. Plugging $%
\chi _{-}=0$ into (\ref{chieq}), one obtains 
\begin{equation}
(g\Sigma _{i}-iX_{I}q^{I}\epsilon _{ij}\Sigma ^{j})\chi _{+}=0.
\label{eqchi+}
\end{equation}
In order to have nontrivial solutions to this equation, the determinant of $%
g\Sigma _{i}-iZ\epsilon _{ij}\Sigma ^{j}$ has to vanish. This implies 
\begin{equation}
Z=\pm g
\end{equation}
for the magnetic central charge $Z$. One can easily show that the lower sign
is incompatible with the condition $\sigma ^{3}\chi _{+}=\chi _{+}$, so the
upper positive sign is chosen and (\ref{eqchi+}) is then automatically
satisfied. The remaining Killing spinor equation for $\chi _{+}$ reads 
\begin{equation}
{\cal D}_{i}\chi _{+}=0,  \label{killspchi}
\end{equation}
with the gauge- and Lorentz-covariant derivative ${\cal D}_{i}$ given by 
\begin{equation}
{\cal D}_{i}=\partial _{i}+\frac{1}{4}\omega _{i}^{\hat{\imath}\hat{\jmath}%
}\Sigma _{\hat{\imath}\hat{\jmath}}-\frac{3i}{2}gV_{I}A_{i}^{I}.
\end{equation}
The integrability condition for (\ref{killspchi}) is 
\begin{equation}
\lbrack {\cal D}_{i},{\cal D}_{j}]\chi _{+}=0,
\end{equation}
or, equivalently, 
\begin{equation}
(\frac{1}{4}R_{ij\hat{k}\hat{l}}\Sigma ^{\hat{k}\hat{l}}-\frac{3i}{2}%
gV_{I}q^{I}\epsilon _{ij})\chi _{+}=0.  \label{integrab}
\end{equation}
Taking into account that $M_{2}$ is of constant curvature, we have 
\begin{equation}
R_{ijkl}=\frac{R}{2}(g_{ik}g_{jl}-g_{il}g_{jk})
\end{equation}
for the Riemann tensor of $M_{2}$. Using this in (\ref{integrab}), one
immediately obtains 
\begin{equation}
R+6gV_{I}q^{I}=0
\end{equation}
for the scalar curvature $R$ of $M_{2}$. From $q^{I}X_{I}=g$ we have 
\begin{equation}
X^{I}=\frac{q^{I}}{g}
\end{equation}
for the moduli, and thus 
\begin{equation}
R=-6gV_{I}q^{I}=-6g^{2}V_{I}X^{I}=-6g^{2}<0
\end{equation}
for the scalar curvature $R$. This means that, to preserve some
supersymmetry, $M_{2}$ must be diffeomorphic to hyperbolic space $H^{2}$ or
to a quotient thereof.\newline
Using the second Cartan equation, one obtains that the spin connection $%
\omega ^{12}$ on $M_{2}$ is related to the vector potential $A^{I}$ by 
\begin{equation}
q^{I}\omega ^{12}=A^{I}\frac{R}{2}.
\end{equation}
Using this, (\ref{killspchi}) reduces to $\partial _{i}\chi _{+}=0$, so that 
$\chi _{+}$ is independent of the coordinates on $M_{2}$. The remaining
equation to solve is the Killing spinor equation on $M_{3}$ for $\eta $,
i.~e. 
\begin{equation}
\left( \partial _{\alpha }+\frac{1}{4}\omega _{\alpha }^{\hat{\alpha}\hat{%
\beta}}\gamma _{\hat{\alpha}\hat{\beta}}+\frac{3}{4}g\gamma _{\alpha
}\right) \eta =0.  \label{killspeta}
\end{equation}
The integrability conditions for (\ref{killspeta}) yield that $M_{3}$ must
be an Einstein space with cosmological constant $\Lambda =-(3g/2)^{2}$. As
we are in three dimensions, this means that $M_{3}$ is also of constant
curvature, i.~e.~a quotient space of $AdS_{3}$. Note that the chirality
condition $\chi =\chi _{+}$ breaks half of supersymmetry. The amount of
supersymmetry preserved by $M_{3}\times M_{2}$ is then determined by the
solutions of (\ref{killspeta}). If $M_{3}=AdS_{3}$, then the whole solution $%
AdS_{3}\times H^{2}$ is half supersymmetric, in agreement with what we found
above. However, we can also choose $M_{3}$ to be e.~g.~the BTZ black hole.
If we take the extremal rotating BTZ black hole, which preserves one half of
the $AdS_{3}$ supersymmetries \cite{henneaux}, then the solution BTZ$%
_{extr}\times H^{2}$ preserves one quarter of the supersymmetries. We would
like to point out that BTZ$\times H^{2}$ compactifications of $D=5$
anti-de~Sitter gravity without gauge fields were previously considered in 
\cite{kiem}. However, as we showed above, due to the relation $Z=g$ these
configurations cannot be supersymmetric unless some gauge fields are turned
on.

\section{Summary and Discussion}

\label{disc}

To sum up, we presented supersymmetric string solutions of gauged $D=5$, $N=2
$ supergravity coupled to abelian vector multiplets. The main result is the
construction of a BPS black string with hyperbolic transverse space,
preserving one quarter of supersymmetry. The curvature of the $H^{2}$ is
supported by a nonvanishing field strength of the vector fields. The black
strings are thus magnetically charged. In the near-horizon limit, their
geometry approaches the half-supersymmetric product space $AdS_{3}\times
H^{2}$, so we encounter supersymmetry enhancement near the horizon. This
behaviour is similar to the case of ungauged supergravity theories. Note
however, that in the ungauged case, usually supersymmetry is fully restored
near the event horizon.\newline
As the near horizon geometry contains an $AdS_{3}$ factor, it should be
possible to use the $AdS_{3}$ asymptotic symmetry algebra \cite{brown} in
order to count the microstates yielding the Bekenstein-Hawking entropy of
the extremal black string. This was done in \cite{strominger} for the BTZ
black hole, and subsequently generalized in \cite{bal,cl1,cl2,cvetic} to
higher-dimensional black holes containing a BTZ factor near the horizon. In
our case, a similar procedure is hindered by the fact that we get the $M=J=0$
BTZ black hole in the near-horizon limit, so using Cardy's formula we would
obtain an incorrect result for the entropy. Similar difficulties have been
encountered in \cite{kaloper}, where a state counting for extremal black
strings in three dimensions was performed. This suggests that a similar
approach to that in \cite{kaloper} must be used in our case, in order to
overcome the above mentioned difficulties.\newline
It is also of interest to investigate the role of the BPS magnetic black
strings in the AdS/CFT correspondence. Note that the $U(1)^{3}$ truncation
of gauged $D=5$, $N=2$ supergravity can be embedded into type IIB
supergravity \cite{cveduff}. This means that our solutions can be lifted to
ten dimensions, with an internal five-sphere that is distorted by the
one-form gauge fields $A_{\mu }^{I}$. That breaks the isometry group $SO(6)$
of the $S^{5}$ down to a smaller subgroup. In the dual CFT, which is ${\cal N%
}=4$ SYM on ${\Bbb R}^{2}$ $\times $ $H^{2}$(or ${\Bbb {}R}\times $ $%
S^{1}\times H^{2}$, if the coordinate $z$ parametrizes a compact space), the 
$S^{5}$ isometry group becomes the R-symmetry. On the CFT side, we are now
dealing with the presence of nonvanishing background $U(1)^{3}$ currents,
which break the global $SO(6)$ R-symmetry. In principle, it should also be
possible to count the microstates giving rise to the black string entropy
using the dual SYM theory on ${\Bbb R}^{2}\times H^{2}$ in the presence of
these global background $U(1)^{3}$ currents.
As the near-horizon geometry is $AdS_3 \times H^2$, the presented magnetic
black string solutions may also have a holographic interpretation in the sense
that the four-dimensional field theory discussed above flows to a
two-dimensional CFT in the IR.
These issues are currently
under investigation \cite{ks}.\newline
Finally, in the present paper, we also investigated $3+2$ product
compactifications, and showed that only if the internal two-manifold is
diffeomorphic to $H^{2}$, some amount of supersymmetry can be preserved. As
an example we found the one quarter supersymmetric BTZ$_{extr}\times H^{2}$
configuration, where BTZ$_{extr}$ denotes the extremal rotating BTZ black
hole. Considering the equations of motion following from (\ref{action}), one
easily sees that one also can have the nonextremal BTZ black hole tensored
by $H^{2}$ as a solution in presence of magnetic gauge fields. Presumably,
these configurations arise as the near-horizon limit of the nonextremal
generalization of (\ref{magstrsol}).

\section*{Acknowledgements}

The authors would like to thank K.~Khuri-Makdisi and A.~Zaffaroni for useful
discussions.


\begin{thebibliography}{99}
\bibitem{ads}  J.~M.~Maldacena, {\em The large $N$ limit of superconformal
field theories and supergravity}, Adv.~Theor.~Math.~Phys.~{\bf 2} (1998)
231; Int.~J.~Theor.~Phys.~{\bf 38} (1999) 1113;\newline
E.~Witten, {\em Anti-de~Sitter space and holography},\newline
Adv.~Theor.~Math.~Phys.~{\bf 2} (1998) 253;\newline
S.~S.~Gubser, I.~R.~Klebanov and A.~M.~Polyakov, {\em Gauge theory
correlators from non-critical string theory}, Phys.~Lett.~{\bf B428} (1998)
105;\newline
O.~Aharony, S.~S.~Gubser, J.~M.~Maldacena, H.~Ooguri and Y.~Oz, {\em Large $%
N $ field theories, string theory and gravity}, hep-th/9905111.

\bibitem{hawkpage}  S.~W.~Hawking and D.~N.~Page, {\em Thermodynamics of
black holes in anti-de~Sitter space}, Commun.~Math.~Phys.~{\bf 87} (1983)
577.

\bibitem{witten}  E.~Witten, {\em Anti-de~Sitter space, thermal phase
transition, and confinement in gauge theories}, Adv.~Theor.~Math.~Phys.~{\bf %
2} (1998) 505.

\bibitem{strominger}  A.~Strominger, {\em Black hole entropy from
near-horizon microstates}, JHEP {\bf 9802} (1998) 009.

\bibitem{brown}  J.~D.~Brown and M.~Henneaux, {\em Central charges in the
canonical realization of asymptotic symmetries: An example from
three-dimensional gravity}, Commun.~Math.~Phys.~{\bf 104} (1986) 207.

\bibitem{romans}  L.~J.~Romans, {\em Supersymmetric, cold and lukewarm black
holes in cosmological Einstein-Maxwell theory}, Nucl.~Phys.~{\bf B383}
(1992) 395.

\bibitem{london}  L.~A.~J.~London, {\em Arbitrary dimensional cosmological
multi-black holes}, Nucl.~Phys.~{\bf B434} (1995) 709.

\bibitem{perry}  A.~Kostelecky and M.~Perry, {\em Solitonic black holes in
gauged $N=2$ supergravity}, Phys.~Lett.~{\bf B371} (1996) 191.

\bibitem{bcs1}  K.~Behrndt, A.~H. Chamseddine and W.~A. Sabra, {\em BPS
black holes in $N=2$ five dimensional AdS supergravity}, Phys.~Lett.~{\bf %
B442} (1998) 97.

\bibitem{ck}  M.~M.~Caldarelli and D.~Klemm, {\em Supersymmetry of
anti-de~Sitter black holes}, Nucl.~Phys.~{\bf B545} (1999) 434.

\bibitem{klemm}  D.~Klemm, {\em BPS black holes in gauged $N=4$, $D=4$
supergravity}, Nucl.~Phys.~{\bf B545} (1999) 461.

\bibitem{bcs2}  K.~Behrndt, M.~Cvetic and W.~A.~Sabra, {\em Non-extreme
black holes of five dimensional $N=2$ AdS supergravity}, Nucl.~Phys.~{\bf %
B553} (1999) 317.

\bibitem{duff}  M.~J.~Duff and J.~T.~Liu, {\em Anti-de~Sitter black holes in
gauged $N=8$ supergravity}, Nucl.~Phys.~{\bf B554} (1999) 237.

\bibitem{liu}  J.~T.~Liu and R.~Minasian, {\em Black holes and membranes in
AdS$_7$}, Phys.~Lett.~{\bf B457} (1999) 39.

\bibitem{sabra}  W.~A.~Sabra, {\em Anti-de~Sitter BPS black holes in $N=2$
gauged supergravity}, Phys.~Lett.~{\bf B458} (1999) 36.

\bibitem{chamsabra}  A.~H.~Chamseddine and W.~A.~Sabra, {\em Magnetic
strings in five dimensional gauged supergravity theories}, hep-th/9911195.

\bibitem{gst2}  M.~G\"{u}naydin, G.~Sierra and P.~K.~Townsend, {\em Gauging
the $D=5$ Maxwell-Einstein supergravity theories: More on Jordan algebras},
Nucl.~Phys.~{\bf B253} (1985) 573.

\bibitem{cy}  P.~S.~Howe and P.~K.~Townsend, {\em Supermembranes and the
modulus space of superstrings}, Talk given at Trieste Conference on
Supermembranes and Physics in $2+1$ Dimensions, Trieste, Italy, Jul 17 - 21,
1989, Published in Trieste Supermembr.~(1989) 165-172.

\bibitem{dewit}  B.~de~Wit and A.~Van Proeyen, {\em Broken sigma-model
isometries in very special geometry}, Phys.~Lett.~{\bf B293} (1992) 94.

\bibitem{feka}  S.~Ferrara and R.~Kallosh, {\em Universality of
supersymmetric attractors}, Phys.~Rev.~{\bf D54} (1996) 1525.

\bibitem{magnetic}  A.~H.~Chamseddine and W.~A.~Sabra, {\em Calabi-Yau black
holes and enhancement of supersymmetry in five dimensions}, Phys.~Lett.~{\bf %
B460} (1999) 63.

\bibitem{birmingh}  D.~Birmingham, {\em Topological black holes in
anti-de~Sitter space}, Class.~Quant.~Grav.~{\bf 16} (1999) 1197.

\bibitem{pope}  H.~L\"{u}, C.~N.~Pope, and J.~Rahmfeld, {\em A construction
of Killing spinors on $S^n$}, hep-th/9805151.

\bibitem{henneaux}  O.~Coussaert and M.~Henneaux, {\em Supersymmetry of the
2+1 black holes}, Phys.~Rev.~Lett.~{\bf 72} (1994) 183.

\bibitem{kiem}  Y.~Kiem and D.~Park, {\em BTZ black holes from the
five-dimensional general relativity with a negative cosmological constant},
Phys.~Lett.~{\bf B450} (1999) 41.

\bibitem{bal}  V.~Balasubramanian and F.~Larsen, {\em Near horizon geometry
and black holes in four dimensions}, Nucl.~Phys.~{\bf B528} (1998) 229.

\bibitem{cl1}  M.~Cveti\v{c} and F.~Larsen, {\em Near horizon geometry of
rotating black holes in five dimensions}, Nucl.~Phys.~{\bf B531} (1998) 239.

\bibitem{cl2}  M.~Cveti\v{c} and F.~Larsen, {\em Microstates of
four-dimensional rotating black holes from near-horizon geometry},
Phys.~Rev.~Lett.~{\bf 82} (1999) 484.

\bibitem{cvetic}  M.~Cveti\v{c}, {\em Microscopics of rotating black holes:
Entropy and greybody factors}, Based on talks given at PASCOS '98, Strings
'98, and Buckow '98, hep-th/9810142.

\bibitem{kaloper}  N.~Kaloper, {\em Entropy count for extremal
three-dimensional black strings}, Phys.~Lett.~{\bf B434} (1998) 285.

\bibitem{cveduff}  M.~Cveti\v{c}, M.~J.~Duff, P.~Hoxha, J.~T.~Liu, H.~Lu,
J.~X.~Lu, R.~Martinez-Acosta, C.~N.~Pope, H.~Sati and T.~A.~Tran, {\em %
Embedding AdS black holes in ten and eleven dimensions}, hep-th/9903214.

\bibitem{ks}  D.~Klemm and W.~A.~Sabra, in preparation.
\end{thebibliography}
\end{document}